\documentclass[11pt]{article}  

\usepackage{graphicx}
\usepackage[all]{xy}
\usepackage{latexsym}
\usepackage[english]{babel}
\usepackage{amsmath}
\usepackage{amsfonts}
\usepackage{amssymb}
\usepackage{makeidx}
\usepackage[utf8]{inputenc}
\usepackage{lmodern}
\usepackage{verbatim}
\usepackage{cite} %must be off to use bibtex
\usepackage[pagewise]{lineno}%\linenumbers
\usepackage{amsthm}
\usepackage{tikz-cd} %Not working on icmat

\newtheorem{theorem}{Theorem}[section]

\newtheorem{lemma}[theorem]{Lemma}

\theoremstyle{definition}

\newtheorem{remark}[theorem]{Remark}
\newtheorem{example}[theorem]{Example}

\newcommand{\R}{\ensuremath{\mathbb{R}}}

\newcommand{\D}{\ensuremath{\mathcal{D}}}
\newcommand{\Es}{\ensuremath{\mathbb{S}}}

\newcommand{\M}{\ensuremath{\mathcal{M}}}

\newcommand{\F}{\ensuremath{\mathbb{F}}}

\usepackage[square, authoryear]{natbib}
\numberwithin{equation}{section}
\numberwithin{figure}{section}

\begin{document}
	\title{Contact bundle formulation of nonholonomic Maupertuis-Jacobi principle and a length minimizing property of nonholonomic dynamics}
	\author{
		{\bf\large Alexandre Anahory Simoes}\hspace{2mm}
		\vspace{1mm}\\
		{\small  Instituto de Ciencias Matem\'aticas (CSIC-UAM-UC3M-UCM)} \\
		{\small C/Nicol\'as Cabrera 13-15, 28049 Madrid, Spain}\\
		{\it\small e-mail: \texttt{alexandre.anahory@icmat.es }}\\
		\vspace{2mm}\\
		{\bf\large Juan Carlos Marrero}\hspace{2mm}
		\vspace{1mm}\\
		{\it\small 	ULL-CSIC Geometr{\'\i}a Diferencial y Mec\'anica Geom\'etrica,}\\
		{\it\small  {Departamento de Matem\'aticas, Estad{\'\i}stica e I O, }}\\
		{\it\small  {Secci\'on de Matem\'aticas, Facultad de Ciencias}}\\
		{\it\small  Universidad de la Laguna, La Laguna, Tenerife, Canary Islands, Spain}\\
		{\it\small e-mail: \texttt{{jcmarrer@ull.edu.es}} }\\
		\vspace{2mm}\\
		{\bf\large David Martín de Diego}\hspace{2mm}
		\vspace{1mm}\\
		{\small  Instituto de Ciencias Matem\'aticas (CSIC-UAM-UC3M-UCM)} \\
		{\small C/Nicol\'as Cabrera 13-15, 28049 Madrid, Spain}\\
		{\it\small e-mail:  \texttt{david.martin@icmat.es} }\\		
	}

	\date{}
	
	\maketitle
	
	\vspace{0.5cm}
	\begin{abstract}
		We prove a nonholonomic version of the classical Mauper\-tuis-Jacobi principle which transforms an autonomous  mechanical nonholonomic problem, determined by a kinetic minus potential  energy and a distribution, in a kinetic nonholonomic problem over a fixed level set of the Lagrangian energy. To prove this result we introduce an appropriate contact bundle structure clarifying the geometric equivalence between both problems. 
		By using the nonholonomic Maupertuis-Jacobi principle, we prove that the regular solutions of a mechanical nonholonomic problem starting from a fixed point and in the same level set of the Lagrangian energy are reparametrizations of geodesics for a family of Riemannian metrics defined on the image of the nonholonomic exponential map. In particular, these trajectories minimize Riemannian length.
	\end{abstract}
	
	%%%%%%%%%%%%%%%%%%%%%%%%%%%%%%%%%%%%%%%%%%%%%%
	\let\thefootnote\relax\footnote{\noindent AMS {\it Mathematics Subject Classification ({2020})}. Primary 70G45; Secondary  53B20, 53C21,
		37J60, 70F25
	}
	%%%%%%%%%%%%%%%%%%%%%%%%%%%%%%%%%%%

\section{Introduction}

One of the most fruitful ideas in mechanics is the intimate relationship between Riemannian geometry and  Lagrangian and Hamiltonian mechanics. 
In particular, a classical and important result in mechanics, the Maupertuis-Jacobi principle (see, for instance, \cite{Arnold,Biesiada})  establishes the relationship between  solutions of a mechanical problem describing motion  in a potential field and the  geodesic motion for a modified Riemannian metric. More precisely, given a Riemannian (or semi-Riemannian) metric  $g$ on a differentiable manifold and $V: Q\rightarrow {\mathbb R}$ a potential function then a mechanical Lagrangian system is determined by the Lagrangian $L_{(g, V)}: TQ\rightarrow {\mathbb  R}$:
\[
L_{(g,V)}(v_q)=\frac{1}{2} g(v_q, v_q)-V(q)\; ,\qquad v_q\in T_qQ\; .
\]
In the region of the configuration space where $V(q)<e$ where $e$ is a constant we define the Riemannian metric (Jacobi metric) by: 
\[
g_e=(e-V)g
\]
Then, it is possible to prove (see \cite{AM78,Godbillon}) that the solutions of the Euler-Lagrange equations for the autonomous mechanical Lagrangian $L_{(g,V)}$ with energy $e$ are the same as the geodesics of the Jacobi metric $g_e$ with energy $1$ up to reparametrization.  This result is known as the Maupertuis-Jacobi principle. 
This important result opens the way to the use of well known techniques in  Riemannian geometry to the study of the qualitative behaviour of the trajectories of mechanical systems as, for instance,  in topological methods to find periodic trajectories of conservative dynamical systems, stability of trajectories, integrability, etc. (see \cite{BKF} and references therein). 

However the problem for nonholonomic mechanics \cite{Bloch,Neimark,Cortes} has been less covered in the literature (see \cite{Koiller}).   One of the reasons is that the equivalent theorem (see Theorem \ref{maup-noh} or   Proposition  8.1 in \cite{Koiller} and \cite{Baksa}) relates a nonholonomic problem given by a mechanical Lagrangian $L_{(g,V)}$ and  a nonintegrable distribution $\D$ with the solutions of a nonholonomic kinetic problem determined by the Lagrangian $L_{g_e}$ and the distribution $\D$. 
It is clear that the new system is now determined by a Riemannian  metric (without potential) but since the motion is nonholonomically constrained by the distribution $\D$ the obtained equations are also not of a variational type and, in principle, it seems impossible to use standard techniques of Riemannian geometry to analyse its qualitative behaviour. 
But, recently, we have shown in   \cite{AMM3} that for kinetic nonholonomic systems,  the nonholonomic solutions starting from a fixed point $q\in Q$  are true geodesics for a family of Riemannian metrics on the image submanifold ${\mathcal M}_q^{nh}$ of the nonholonomic exponential map at $q$. 
Therefore, by using these special Riemannian metrics combined with the construction of the Jacobi metric for nonholonomic systems, we can deduce that regular nonholonomic mechanical trajectories starting from a fixed point $q\in Q$ and in the same level set of the Lagrangian energy are reparametrizations of true geodesics for a family of Riemannian metrics on the image submanifold of the nonholonomic exponential map. In particular, these nonholonomic trajectories minimize the Riemannian length for sufficiently small times. This is the main result of the paper. Note that, as a direct consequence,  we could use Riemannian tools to analyse the qualitative behaviour of nonholonomic systems of the type $(L_{(g, V)}, \D)$.

The paper is structured as follows. 
In Section \ref{section1}, we introduce the equations of motion of a kinetic nonholonomic system using tools of Riemannian geometry (see \cite{Lewis98, Synge28}). Then, we recall the definition of the nonholonomic exponential map introduced in \cite{AMM, AMM2} and its main properties. In addition, section \ref{section1} also contains a result, Theorem \ref{kin-theorem} (proved in \cite{AMM3}), which shows that kinetic nonholonomic trajectories starting from the same initial point can be seen as geodesics of an appropriate family of Riemannian metrics .
In Section \ref{section3} we move on to analyse the case of a mechanical nonholonomic system where we now consider an additional potential energy and we construct an associated kinetic nonholonomic problem introducing the associated Jacobi metric. With these elements, we state the main result of the paper, Theorem  \ref{mechanical:nonholonomic:geodesic:theorem}, showing the minimizing property of regular nonholonomic solutions on level sets of the Lagrangian energy. To give a complete proof of this Theorem we develop in the next sections all the necessary mathematical tools. In particular, in Theorem \ref{maup-noh} (Section  \ref{section4}), we prove a contact bundle formulation of the Maupertuis-Jacobi principle. For that purpose, we will use the symplectic bundle formulation of mechanical nonholonomic systems which was proposed in \cite{BaSn}. With Theorem \ref{maup-noh} and 
Theorem \ref{kin-theorem}, the proof of Theorem \ref{mechanical:nonholonomic:geodesic:theorem} follows as a corollary.

\section{Nonholonomic exponential map for kinetic nonholonomic systems}\label{section1}
In this section, we will review the definition of the nonholonomic exponential map for a kinetic nonholonomic system and some results on this map (for the definition of the nonholonomic exponential map associated with an arbitrary nonholonomic system and its properties, see \citep*{AMM}).

First of all, we will see that the solutions of the equations of motion of a kinetic nonholonomic system are the geodesics of a  constrained connection (the nonholonomic connection) on the configuration space restricted to initial conditions in $\D$ \cite*{Lewis98}. This construction seems to have been first made in \citep*{Synge28}.

As we have commented in the introduction, a kinetic nonholonomic system is determined by a triple $(Q, g, \D)$, where $Q$ is a finite dimensional smooth manifold, $g$ is a Riemannian metric on $Q$ and $\D$ is a nonintegrable distribution determining the nonholonomic constraints \cite*{LMdD1996}. 

The nonholonomic connection $\nabla^{nh}$ is   defined as
\begin{equation}\label{nhconnection}
\nabla^{nh}_{X} Y:=P(\nabla_{X}^{g} Y)+\nabla^{g}_{X}[P'(Y)], \; \; \mbox{ for } X, Y \in \frak{X}(Q),
\end{equation}
where  $P:TQ\rightarrow \D$ is the associated  orthogonal projector onto the distribution $\D$ and $P':TQ\rightarrow\D^{\bot}$ is the orthogonal projector onto $\D^{\bot}$, the orthogonal distribution.

This connection is not symmetric (that is, it is not torsion free, see \cite{BLMMM2011}, for an alternative symmetry condition) and  in general neither it is compatible with the metric. Nevertheless, it satisfies a more restricted condition of compatibility with the Riemannian metric $g$ over sections of $\D$ (see \citep*{Lewis98}), i.e.,
\begin{equation}\label{Dcompatibility}
X(g(Y,Z))=g(\nabla^{nh}_{X} Y,Z)+g(Y,\nabla^{nh}_{X} Z), \quad \forall X,Y,Z\in\Gamma(\D).
\end{equation}
It is interesting to note that if   $Y\in\Gamma(\D)$ then $\nabla^{nh}_{X} Y=P(\nabla_{X}^{g} Y)\in \Gamma(\D)$ for any vector field $X\in\mathfrak{X}(Q)$.
%Observe that Equation (\ref{Dcompatibility}) is equivalent to 

The  geodesics $c$ for this connection which satisfy the constraints, that is, 
\begin{equation}\label{cnh-1}
\nabla^{nh}_{\dot{c}(t)}\dot{c}(t)=0\; ,\qquad   \dot{c}(0)\in \D_{c(0)}
\end{equation}
are precisely the solutions of the nonholonomic problem given by $(Q, g, \D)$ (see, for instance, \citep*{Lewis98, BLMMM2011}). %Observe that this equation is equivalent to Equations (\ref{cnh}).

\begin{lemma}[\cite{AMM3}]\label{homogeneity}
{Let $c_v: I \to Q$ be a nonholonomic geodesic with initial velocity $v \in \D_q$, i.e.}
\[
c_{v}(t_0) = q \; \; \mbox{ and } \; \; \dot{c}_{v}(t_0) = v.
\]
\begin{enumerate}
\item
{We have that
\begin{equation}\label{constant-norm}
\|\dot{c}_v(t) \|_{g(c_v(t))} = \| v\|_{g(q)}, \; \; \mbox{ for } t \in I.
\end{equation}}
\item
If $v = 0$ then $c_v(t) = q$, for every $t \in I$.
\item
If $v \neq 0$ then a reparametrization of $c_v$,
\[
c_v \circ r: J \to Q, \; \; s \to c_v(r(s))
\]
is a nonholonomic geodesic if and only if
\[
r(s) = as + b, \; \; \mbox{ with } a, b \in \mathbb{R}.
\]
\end{enumerate}
\end{lemma}
The tangent lifts of the nonholonomic geodesics of a kinetic nonholonomic system $(Q, g, \D)$ are the integral curves of a vector field of  $\Gamma_{(g,\D)}\in {\mathfrak X}(\D)$, which is a second-order differential equation along the points of $\D$, considered as a vector subbundle of $TQ$ (see, for instance, \citep*{LMdD1996}). 

Denote by $\phi_t^{\Gamma_{(g,\D)}}: \D \rightarrow \D$ the flow of $\Gamma_{(g,\D)}$ and for a sufficiently small positive number $h$, we consider the open subset of $\D$ given by
\begin{equation*}
M_{h}^{\Gamma_{(g,\D)}}=\{ v\in\D \ | \ \phi_{t}^{\Gamma_{(g,\D)}}(v) \ \text{is defined for} \ t\in [0,h] \}.
\end{equation*}

Using the last part of Lemma \ref{homogeneity} we can assume, without the loss of generality, that $h=1$. Then, we will denote the open subset 
$M_{1}^{\Gamma_{(g,\D)}}$ of $\D$ by $M^{\Gamma_{(g,\D)}}$. {In addition, from the second part of Lemma \ref{homogeneity}, we also have that the zero section in $\D$ is contained in $M^{\Gamma_{(g,\D)}}$.}

From the flow of $\Gamma_{(g,\D)}$, we can define the nonholonomic exponential map
	\begin{align*}
	\text{exp}^{\Gamma_{(g,\D)}}:M^{\Gamma_{(g,\D)}}\subseteq \D & \rightarrow  Q \times Q\\
	v & \mapsto(\tau_Q(v), \tau_{Q}\circ\phi_1^{\Gamma_{(g,\D)}}(v))
	\end{align*}
(see \citep*{AMM}).	
	We remark that if $c_{v}: [0, 1]\rightarrow Q$ is the nonholonomic geodesic with $\dot{c}_{v}(0)=v$ then
		\begin{equation}\label{nh-exponential-map}
		\text{exp}^{\Gamma_{(g,\D)}}(v)=(\tau_Q(v), c_{v}(1)).
	         \end{equation}
	
	We will use in the sequel the restriction of this map to the open subset $M_{q}^{\Gamma_{(g,\D)}} = M^{\Gamma_{(g,\D)}} \cap \D_{q} $ of $\D_{q}$ with $q\in Q$ fixed, that is, we define
	\[
	\text{exp}_{q}^{nh}=	\text{pr}_2 \circ \text{exp}^{\Gamma_{(g,\D)}}\Big|_{M_{q}^{\Gamma_{(g,\D)}}}: M_{q}^{\Gamma_{(g,\D)}} \subset \D_{q}\longrightarrow Q
	\]
So, if $v_q \in \D_{q}$ and $c_{v_q}: [0, 1] \to Q$ is the nonholonomic geodesic with initial velocity $v_q$ then
\[
\text{exp}_{q}^{nh}(v_q) = c_{v_q}(1).
\]
The reader is invited to compare the definition of $\text{exp}_{q}^{nh}$ with that of the Riemannian exponential at $q$ (see (\cite{docarmo}).

In fact, the nonholonomic exponential map conserves many of the properties we may find in Riemannian exponential maps.

The most important result is the theorem showing that kinetical nonholonomic trajectories starting form the same initial point can be seen as geodesics of an appropriate family of Riemannian metrics (see \cite{AMM3}). 

\begin{theorem}\label{kin-theorem}
	Let $(Q, g,\D)$ be a kinetic nonholonomic system and $q$ a fixed point in $Q$.Then:
	\begin{enumerate}
		\item[i)]
		There exists a submanifold ${\mathcal M}^{nh}_{q}$ of $Q$, with  $q\in {\mathcal M}^{nh}_{q}$, and a diffeomorphism
		$\emph{exp}^{nh}_{q}: {\mathcal U}_0 \subseteq \D_{q} \to {\mathcal M}^{nh}_{q} \subseteq Q$,
		{where $\, {\mathcal U}_0$ is a starshaped open subset of $\D_{q}$ about $0_{q}\in {\mathcal U}_{0}$} and $\emph{exp}^{nh}_{q}(0_{q}) = q$. The map $\emph{exp}^{nh}_{q}$ is the nonholonomic exponential map at $q$. Moreover, we have that:
		\begin{enumerate}
			\item
			Under the canonical linear identification between $\D_{q}$ and $T_{0_{q}}{\mathcal U}_0$, the linear monomorphism
			\[
			T_{0_{q}}\emph{exp}^{nh}_{q}: T_{0_{q}}{\mathcal U}_0 \simeq \D_{q} \to T_{q}Q
			\]
			is just the canonical inclusion of $\D_{q}$ in $T_{q}Q$.
			\item
			For every $v_{q} \in {\mathcal U}_{0}$,
			\begin{equation}\label{radial:standard:form}
			\emph{exp}^{nh}_{q}(tv_{q}) = c_{v_{q}}(t), \; \; t \in [0, 1],
			\end{equation}
			with $c_{v_{q}}: [0, 1] \to {\mathcal M}^{nh}_{q} \subseteq Q$ the (unique) nonholonomic trajectory satisfying
			$c_{v_{q}}(0) = q, \dot{c}_{v_{q}}(0) = v_{q}$.
		\end{enumerate}   
		\item[ii)] All the radial kinetic nonholonomic trajectories departing from the fixed point $q\in Q$ are {homothetic reparametrizations} of nonholonomic trajectories {given by equation \eqref{radial:standard:form}}. In addition, they are minimizing geodesics for a Riemannian metric $g_{q}^{nh}$ on ${\mathcal M}^{nh}_{q}$ if and if only if the Riemannian metric ${\mathcal G}_0 = (\emph{exp}^{nh}_{q})^*(g_{q}^{nh})$ on ${\mathcal U}_0$ satisfies the Gauss condition, that is,
		\[
		{\mathcal G}_{0}(v_{q})(v_{q}, w_{q}) = {\mathcal G}_{0}(0_{q})(v_{q}, w_{q}), \; \; \mbox{ for } v_{q} \in {\mathcal U}_0 \mbox{ and } w_{q} \in D_{q}.
		\]
		\item[iii)] Such Riemannian metrics on ${\mathcal M}^{nh}_{q}$ always exist and if $g^{nh}_q$ is one of them then the Riemannian exponential associated with $g^{nh}_q$ at $q$ is just $\emph{exp}^{nh}_q$.		
	\end{enumerate}
\end{theorem}

\section{Mechanical nonholonomic trajectories}\label{section3}

In this section, we state a nonholonomic version of the Maupertuis-Jacobi  principle.

Then, using Theorem \ref{kin-theorem}, we will immediately deduce that radial nonholonomic mechanical trajectories with fixed energy $e\in\R$ are, for sufficiently small times, strictly increasing reparametrizations of minimizing Riemannian geodesics on a suitable Riemannian manifold.

Let $g$ be a Riemannian metric on  the $n$-dimensional manifold $Q$, $V:Q\rightarrow \R$ be a smooth function called the potential energy and let $\D$ be a rank $r$ distribution on $Q$. Let $L_{(g,V)}:TQ\rightarrow \R$ be the mechanical Lagrangian function associated with the Riemannian metric $g$ and potential energy $V$, that is,
\begin{equation*}
	L_{(g,V)}(v)=\frac{1}{2}g(v,v)-V\circ \tau_{Q}(v), \quad v\in TQ.
\end{equation*}
The triple $(Q, {(g,V)},\D)$ is called a \textit{nonholonomic mechanical system} \cite*{Bloch}. The trajectories of a nonholonomic mechanical system satisfy the equations: 
\begin{equation}\label{cnh-2}
\nabla^{nh}_{\dot{c}(t)}\dot{c}(t)+\hbox{grad}^g V(c(t))=0\; ,\qquad   \dot{c}(0)\in \D_{c(0)},
\end{equation}
where $\hbox{grad}^g$ is the gradient vector field on $Q$ associated with the potential energy $V$ via the metric $g$, that is,
\begin{equation*}
	g(\hbox{grad}^g V,X) = \langle dV,X \rangle, \quad \forall X \in \mathfrak{X}(Q).
\end{equation*}
Therefore, given $v_{q}\in \D$, denote by $c_{v_{q}}:I\rightarrow Q$ the unique solution of (\ref{cnh-2}) with initial conditions
\begin{equation*}
	c_{v_{q}}(0)=q, \quad \dot{c}_{v_{q}}(0)=v_{q}.
\end{equation*}
As it is well-known, $(Q,{(g,V)},\D)$ is a regular nonholonomic mechanical system and thus the tangent lift of the trajectories $c_{v_{q}}$ (which we denote by $\dot{c}_{v_{q}}:I\rightarrow TQ$) are integral curves of a SODE denoted by $\Gamma_{(g,V,\D)}\in\mathfrak{X}(\D)$.

The energy of the system $(Q, L_{(g,V)}, \D)$ is given by the function $E_{{(g,V)}}:\D\rightarrow \R$ defined by
\begin{equation*}
	E_{{(g,V)}}(v)=\frac{1}{2}g(v,v)+V\circ \tau_{Q}(v), \quad v\in \D.
\end{equation*}
Recall that the energy is a first integral of the vector field $\Gamma_{(g,V,\D)}$, which implies that the energy is constant along the trajectories $c_{v_{q}}$, i.e.,
\begin{equation*}
	E_{{(g,V)}}(\dot{c}_{v_{q}}(t))=e, \quad \forall t\in I,
\end{equation*}
where $e\in \R$ is some real number. Note that, for mechanical systems $e\geqslant V(c_{v_{q}}(t))$.

Fixing a real number $e\in\R$, it is possible to classify the mechanical trajectories into two different types:
\begin{enumerate}
	\item[(i)] \textbf{Singular trajectories}: the energy of the trajectory $c_{v_{q}}$ satisfies $e=V(q)$, which automatically implies that the initial velocity is zero $v_{q}=0$.
	\item[(ii)] \textbf{Regular trajectories}: the energy of the trajectory $c_{v_{q}}$ satisfies $e>V(q)$ and the velocity of the trajectory may be written as
	\begin{equation*}
		\|\dot{c}_{v_{q}}(t)\|^2=2(e-V(c_{v_{q}}(t))), \quad \forall t\in I.
	\end{equation*}
	So, there exists a real number $\varepsilon>0$ such that the curve $c_{v_{q}}:(-\varepsilon,\varepsilon)\rightarrow Q$ is a regular trajectory.
\end{enumerate}

Now, if for a fixed $e\in\R$, the curve $c_{v_{q}}$ is a regular trajectory, that, is $e>V(q)$, then it is clear that the initial velocity is in the sphere centered at the zero vector $0_{q}$ and with radius $\sqrt{2(e-V(q))}$, which we will denote by
\begin{equation*}
	v_{q}\in S_{g}\left( \sqrt{2(e-V(q))} \right),
\end{equation*}
where the subscript $g$ indicates that the norm is measured relative to the Riemannian metric $g$.

\begin{remark}
{\rm
The set $ 
\{q\in Q\; |\; e\geq V(q)\}
$ is usually called the Hill region and the set $ 
\{q\in Q\; |\; e = V(q)\}$ is called the Hill boundary or also sometimes called the zero velocity surface.
 }
\end{remark}

Now, take $e\in\R$ such that
\begin{equation*}
	U_{e}=\{ q\in Q \ | \ e>V(q) \}
\end{equation*}
is a non-empty subset of $Q$. Then, $U_{e}$ is an open subset of $Q$ and, if it is non-empty, it inherits the smooth manifold structure of $Q$. We can consider on it the \textit{Jacobi metric}
\begin{equation}\label{jacmet}
	g_{e}=(e-V)g
\end{equation}
and the kinetic nonholonomic system $(U_{e},g_{e},\D_{e})$, where the distribution $\D_{e}$ is nothing but the fibers of $\D$ at the points in $U_{e}$. In other words, $\D_{e}=(\tau_{\D})^{-1}(U_{e})$, where $\tau_{\D}:\D\rightarrow Q$ is the vector bundle projection.

Given a vector $v_{q}\in \D_{e}$, we will denote by $c_{v_{q}}^{e}:I\rightarrow U_{e}$ the nonholonomic trajectory of $(U_{e},g_{e},\D_{e})$, with initial velocity $v_{q}$, that is
\begin{equation}\label{cnh-e}
\nabla^{nh, e}_{\dot{c}_{v_{q}}^{e}(t)}\dot{c}_{v_{q}}^{e}(t)=0\; ,\qquad   \dot{c}_{v_{q}}^{e}(0)\in \D_{c_{v_{q}}^{e}(0)}
\end{equation}
where $\nabla^{nh,e}_{X} Y:=P(\nabla_{X}^{g_e} Y)+\nabla^{g_e}_{X}[P'(Y)],$  $X, Y \in \Gamma (\D_e)$. Observe that since $g_e$ and $g$ are  in the same conformal class of metrics the orthogonal projectors are the same for both metrics. 

Therefore, there exists a SODE $\Gamma_{(g_{e},\D_{e})}\in\mathfrak{X}(\D_{e})$ whose integral curve with initial velocity $v_{q}$ is precisely the tangent lift of the trajectory $c_{v_{q}}^{e}$.

Moreover, the energy of this system is simply given by the Lagrangian itself, that is, $E_{L_{g_{e}}}:TU_{e}\rightarrow \R$ coincides with the Lagrangian function $L_{g_{e}}:TU_{e}\rightarrow \R$ given by
\begin{equation*}
	L_{g_{e}}(u)=\frac{1}{2}g_{e}(u,u), \quad u\in TU_{e}.
\end{equation*}
Thus  , $L_{g_{e}}|_{\D_{e}}$ is a first integral of $\Gamma_{(g_{e},\D_{e})}$.

Moreover, it is not difficult to prove that if the trajectory $c_{v_{q}}^{e}$ has energy equal to $1$, then the initial velocity $v_{q}$ satisfies
\begin{equation*}
	v_{q}\in S_{g}\left( \sqrt{\frac{2}{e-V(q)}} \right),
\end{equation*}
using the same notation as before.

Now, let us introduce two projections and a suitable diffeomorphism between the two spheres mentioned above.

Let ${\mathcal P}_{q}:T_{q}Q\setminus \{0_{q}\}\rightarrow S_{g}\left( \sqrt{2(e-V(q))} \right)$ be the projection given by
\begin{equation*}
	{\mathcal P}_{q}(v_{q})=\sqrt{2(e-V(q))} \frac{v_{q}}{\|v_{q}\|_{g}}
\end{equation*}
and ${\mathcal  Q}_{q}:T_{q}Q\setminus \{0_{q}\}\rightarrow S_{g}\left( \sqrt{\frac{2}{e-V(q)}} \right)$ be the projection given by
\begin{equation*}
	{\mathcal  Q}_{q}(v_{q})=\sqrt{\frac{2}{e-V(q)}} \frac{v_{q}}{\|v_{q}\|_{g}}.
\end{equation*}
Consider the map $\Psi_{q}:S_{g}\left( \sqrt{2(e-V(q))} \right) \rightarrow S_{g}\left( \sqrt{\frac{2}{e-V(q)}} \right)$ that makes the  diagram of Figure  \ref{psi1} to commute.
\begin{figure}[h]
	\begin{center}
	\begin{tikzcd}
		& T_{q}Q\setminus\{0_{q}\} \arrow[ld, "{\mathcal P}_{q}"'] \arrow[rd, "\mathcal{Q}_{q}"] &                                             \\
		S_{g}\left( \sqrt{2(e-V(q))} \right) \arrow[rr, "\Psi_{q}"] &                                                                             & S_{g}\left( \sqrt{\frac{2}{e-V(q)}} \right)
	\end{tikzcd}
	\end{center}
\caption{Definition of the diffeomorphism $\Psi_q$ between spheres}\label{psi1}
\end{figure}
Observe that $\Psi_{q}$ is a diffeomorphism with  explicit expression
\begin{equation*}
	\Psi_{q}(v_{q})=\frac{1}{(e-V(q))}v_{q}.
\end{equation*}

We are now in position to formulate the  main result of this paper.

\begin{theorem}\label{mechanical:nonholonomic:geodesic:theorem}
	Let $(Q,{(g,V)},\D)$ be a mechanical nonholonomic system, $q \in Q$ a fixed point of the manifold $Q$ and let $e\in \R$ such that $e>V(q)$. Then:
	\begin{enumerate}
		\item[i)] There exists $\varepsilon>0$ and a submanifold $\M_{q}^{nh,e}\subset Q$ with $q \in \M_{q}^{nh,e}$ and a diffeomorphism
		\begin{equation*}
			\emph{exp}_{q}^{nh,e}:B_{g}\left( 0_{q};\sqrt{\frac{{2 \varepsilon}}{e-V(q)}} \right)\subseteq \D_{q} \rightarrow \M_{q}^{nh,e},
		\end{equation*}
		where the domain denotes the open ball in $\D_{q}$ around $0_{q}$ with radius $\sqrt{\frac{{2 \varepsilon}}{e-V(q)}}$, with respect to the Riemannian metric $g$.
		
		Moreover we have that $\emph{exp}_{q}^{nh,e}(0_{q})=q$ and:
		\begin{enumerate}
			\item[(a)] The tangent map of $\emph{exp}_{q}^{nh,e}$ at $0_{q}$, under the canonical linear identification between $\D_{q}$ and $T_{0_{q}}\left(B_{g}\left( 0_{q};\sqrt{\frac{{2 \varepsilon}}{e-V(q)}} \right)\right)$,
			\begin{equation*}
				T_{0_{q}}\emph{exp}_{q}^{nh,e}:\D_{q}\longrightarrow T_{q}Q,
			\end{equation*}
			is just the canonical inclusion of $\D_{q}$ in $T_{q}Q$.
			\item[(b)] For every non-zero vector $v_{q}\in B_{g}\left( 0_{q};\sqrt{\frac{{2 \varepsilon}}{e-V(q)}} \right)$ the nonholonomic mechanical trajectory $c_{{\mathcal P}_{q}(v_{q})}:[0,\lambda]\rightarrow Q$ satisfies
			\begin{equation}\label{radial:mech:trajectories}
				c_{{\mathcal P}_{q}(v_{q})}(s)=\emph{exp}_{q}^{nh,e}(h(s)\mathcal{Q}_{q}(v_{q})),
			\end{equation}
			where $h:[0,\lambda]\rightarrow [0,\delta]$ is a strictly increasing reparametrization satisfying
			\begin{equation*}
				\frac{d h}{d s}=e-V\circ c_{{\mathcal P}_{q}(v_{q})}, \quad h(0)=0
			\end{equation*}
			and $\lambda$ is sufficiently small in such a way that
			\begin{equation*}
				h(s)\mathcal{Q}_{q}(v_{q})\in B_{g}\left( 0_{q};\sqrt{\frac{{2 \varepsilon}}{e-V(q)}} \right), \quad \forall s \in [0,\lambda].
			\end{equation*}
		\end{enumerate}
		\item[ii)] All the nonholonomic trajectories with starting point $q$ and energy $e$ are, for sufficiently small times, of the form \eqref{radial:mech:trajectories}. In addition, if $g_{q}^{nh,e}$ is a Riemannian metric on $\M_{q}^{nh,e}$ such that $\mathcal{G}_{0}^{e}=(\emph{exp}_{q}^{nh,e})^{*}g_{q}^{nh,e}$ satisfies the Gauss condition, then the curves
		\begin{equation*}
			t\in [0,1]\mapsto \emph{exp}_{q}^{nh,e}(tv_{q})\in \M_{q}^{nh,e},
		\end{equation*}
		with $v_{q}\in B_{g}\left( 0_{q};\sqrt{\frac{{2 \varepsilon}}{e-V(q)}} \right)$ are geodesics for $g_{q}^{nh,e}$ and, therefore, the nonholonomic trajectories
		\begin{equation*}
			s\in [0,\lambda] \mapsto c_{P_{q}(v_{q})}(s)\in \M_{q}^{nh,e}
		\end{equation*}
		are reparametrizations of minimizing geodesics for the metric $g_{q}^{nh,e}$. In particular, these nonholonomic trajectories minimize length in $\M_{q}^{nh,e}$.
		\item[iii)] The Riemannian metrics $g_{q}^{nh,e}$ on $\M_{q}^{nh,e}$ always exist.
	\end{enumerate}
\end{theorem}

\begin{remark}
	We have that the map
	\begin{equation*}
		\text{exp}_{q}^{nh,e}:B_{g}\left( 0_{q};\sqrt{\frac{{2 \varepsilon}}{e-V(q)}} \right)\subseteq \D_{q} \longrightarrow U_{e}\subseteq Q
	\end{equation*}
	is given by
	\begin{equation*}
		\text{exp}_{q}^{nh,e}(v_{q})=\tau_{Q}\left( \phi_{1}^{\Gamma_{(g_{e},\D_{e})}} (v_{q})\right)
	\end{equation*}
	for $v_{q}\in B_{g}\left( 0_{q};\sqrt{\frac{{2 \varepsilon}}{e-V(q)}} \right)$ and where $\phi_{t}^{\Gamma_{(g_{e},\D_{e})}}$ is the flow of the SODE $\Gamma_{(g_{e},\D_{e})}$ along $\D_{e}$. In other words, $\text{exp}_{q}^{nh,e}$ is the nonholonomic exponential map at $q$ associated with the kinetic non-holonomic system $(U_{e},g_{e},\D_{e})$.
\end{remark}
\begin{remark}
In Theorem	\ref{mechanical:nonholonomic:geodesic:theorem} (item ii)) we mention the notion of a metric on $\D_q$ satisfying the Gauss condition. This type of metrics were introduced in \cite*{AMM3}. In fact, a Riemannian metric ${\mathcal G}_0$ on a finite-dimensional real vector space $E$ is a Gauss metric if
\[
{\mathcal G}_0 (u) (u, v)={\mathcal G}(0)(u,v), \qquad \forall u, v\in E\; .
\]	
\end{remark}

In order to prove our main theorem we will need the following version of the nonholonomic Maupertuis-Jacobi principle relating nonholonomic mechanical trajectories with nonholonomic trajectories of an associated kinetical nonholonomic problem. 

\begin{theorem}[Nonholonomic Maupertuis-Jacobi  theorem]\label{nhm}
	Let $(Q,{(g,V)},\D)$ be a mechanical nonholonomic system, $q \in Q$ a fixed point of the manifold and let $e\in \R$ such that $e>V(q)$. For a non-zero $v_{q}\in T_{q}U_{e}$ denote by
	\begin{equation*}
		c_{{\mathcal P}_{q}(v_{q})}:J\longrightarrow U_{e} \quad \text{and} \quad c_{\mathcal{Q}_{q}(v_{q})}:I\longrightarrow U_{e} \ \text{with} \ 0\in I,J
	\end{equation*}
	the nonholonomic trajectories for the systems $(U_{e},L_{(g,V)}|_{TU_{e}},\D_{e})$ and $(U_{e},L_{g_{e}},\D_{e})$ with initial velocities $P_{q}(v_{q})$ and $\mathcal{Q}_{q}(v_{q})$, respectively. Then, we have that
	\begin{equation*}
		c_{P_{q}(v_{q})}(s)=c_{\mathcal{Q}_{q}(v_{q})}(h(s)),
	\end{equation*}
	where $h:J\rightarrow I$ is a strictly increasing reparametrization satisfying
	\begin{equation*}
		\frac{d h}{d s}=e-V\circ c_{P_{q}(v_{q})}, \quad h(0)=0.
	\end{equation*}
\end{theorem}

\section{Nonholonomic Maupertuis-Jacobi principle}\label{section4}

In this section, we develop the machinery we will need to prove the nonholonomic Maupertuis-Jacobi  Theorem \ref{nhm}.

\subsection{Symplectic bundle formulation of nonholonomic mechanical systems}

Let $(Q,{(g,V)},\D)$ be a mechanical nonholonomic system with $\text{rank}\, \D=r$.

\subsubsection{The Lagrangian side}\label{lasub}

We will review the main ingredients of the construction given by  \cite{BaSn} (see also \cite{Cortes, CdLMM2009}).
First of all, we will introduce the set
\begin{equation*}
	\mathcal{T}^{\D}\D = \bigcup\limits_{\begin{split}
		v_{q} & \in\D_{q}\\ q & \in Q
		\end{split}} \{ X\in T_{v_{q}}\D \ | \ (T_{v_{q}}\tau_{Q})(X)\in \D_{q}  \}
\end{equation*}
which is  a symplectic vector bundle of rank $2 r$ over $\D$, that is, $$(\mathcal{T}^{\D}_{v_{q}}\D,\left.\omega_{L_{(g,V)}}(v_{q})\right|_{\mathcal{T}^{\D}_{v_{q}}\D})$$ is a symplectic vector space of dimension $2r$, for all $v_{q}\in\D_{q}$, where $\omega_{L_{(g,V)}}$ is the Poincaré-Cartan $2$-form associated with the mechanical Lagrangian $L_{(g,V)}$ (see \cite{LR89}).

Let $E_{(g,V)}$ be the corresponding Lagrangian energy. Then we have that
\begin{equation*}
	\left. d E_{(g,V)}(v_{q}) \right|_{\mathcal{T}^{\D}_{v_{q}}\D} \in (\mathcal{T}^{\D}_{v_{q}}\D)^{*}, \hbox{   for all   }  v_q\in \D_q   .
\end{equation*}
Moreover, we have that the nonholonomic vector field $\Gamma_{(g,V,\D)}$ defined in (\ref{cnh-2}) is geometrically characterized by the equations
\begin{equation}\label{geometric:nh:symp:eq}
	\begin{split}
		& \left( i_{\Gamma_{(g,V,\D)}} \omega_{L_{(g,V)}}|_{\D} \right)|_{\mathcal{T}^{\D}\D}=\left( d E_{(g,V)}|_{\D} \right)|_{\mathcal{T}^{\D}\D} \\
		& \Gamma_{(g,V,\D)}\in \Gamma(\mathcal{T}^{\D}\D).
	\end{split}
\end{equation}
%In fact, equations \eqref{geometric:nh:symp:eq} characterize the vector field %$\Gamma_{(g,V,\D)}$. 
As an immediate consequence, we deduce the preservation of energy for the nonholonomic trajectories: 
\begin{equation}
	\Gamma_{(g,V,\D)}(E_{(g,V)}|_{\D})=0.
\end{equation}

\subsubsection{The Hamiltonian side}

Given a Riemannian metric $g$ and a potential energy function $V$ on the manifold $Q$, we may consider the Hamiltonian function $H_{(g,V)}:T^{*}Q\rightarrow \R$ given by
\begin{equation*}
	H_{(g,V)}(\alpha_{q})=\frac{1}{2}g_{q}^{\sharp}(\alpha_{q},\alpha_{q})+V(q), \quad \alpha_{q}\in T_{q}^{*}Q,
\end{equation*}
where we are denoting by $g^{\sharp}$ the co-metric associated to the Riemannian metric $g$. Indeed, given a Riemannian metric $g$, there is an isomorphism of modules $\flat_{g}:\mathfrak{X}(Q)\rightarrow \Omega^{1}(Q)$ called the \textit{flat isomorphism} given by
\begin{equation*}
	\langle \flat_{g}(X(q)), Y(q) \rangle=g_{q}(X(q),Y(q)), \quad X,Y\in\mathfrak{X}(Q).
\end{equation*}
Then the co-metric is the map $g^{\sharp}:\Omega^{1}(Q)\times \Omega^{1}(Q)\rightarrow C^\infty(Q)$ given by
\begin{equation*}
	g^{\sharp}_{q}(\flat_{g}(X(q)),\flat_{g}(Y(q)))=g_{q}(X(q),Y(q)), \quad X,Y\in\mathfrak{X}(Q).
\end{equation*}

It is also interesting to note that the Legendre transform of the mechanical Lagrangian function $L_{(g,V)}$, denoted by $\F L_{(g,V)}:TQ\rightarrow T^{*}Q$, coincides with the flat isomorphism, i.e.,
\begin{equation*}
	\F L_{(g,V)}=\flat_{g}\; .
\end{equation*}
Moreover, we have that
\begin{equation*}\label{omeh}
	(\F L_{(g,V)})^{*}\omega_{Q}=\omega_{L_{(g,V)}} \quad \text{and} \quad (\F L_{(g,V)})^{*}H_{(g,V)}=E_{{(g,V)}}
\end{equation*}
where $\omega_Q$ is the canonical symplectic form on $T^*Q$. 

If $\D^{\bot}$ is the orthogonal complement of $\D$ with respect to the metric $g$ and
\begin{equation*}
	(\D^{\bot})^{o}=\bigcup_{q\in Q} \{ \alpha_{q}\in T_{q}^{*}Q \ | \ \langle \alpha_{q},v_{q} \rangle=0, \ \forall v_{q}\in \D^{\bot}_{q} \}
\end{equation*}
then we have that
\begin{equation*}
	\F L_{(g,V)}(\D)=(\D^{\bot})^{o}.
\end{equation*}
It is clear that $(i^*_{\D})|_{(\D^{\bot})^{o}}: (\D^{\bot})^{o}\rightarrow \D^{*}$ is an isomorphism of vector bundles where $i_{\D}:\D\hookrightarrow TQ$ is the canonical inclusion. From now on, we will use the previous canonical identification between $(\D^{\bot})^{o}$ and $\D^{*}$. We have that $T\F L_{(g,V)}=T\flat_{g}$ is a vector bundle isomorphism over $\F L_{(g,V)}=\flat_{g}$. Hence, considering the following vector bundle over $\D^{*}$
\begin{equation*}
	\mathcal{T}^{\D^{*}}\D^{*} = \bigcup\limits_{\begin{split}
	\alpha_{q} & \in\D_{q}^{*}\\ q & \in Q
	\end{split}} \{ Y\in T_{\alpha_{q}}\D^{*} \ | \ (T_{\alpha_{q}}\pi_{Q})(Y)\in \D_{q}  \},
\end{equation*}
where $\pi_{Q}:T^{*}Q\rightarrow Q$ is the cotangent bundle projection, we have that
\begin{equation*}
	T\flat_{g}(\mathcal{T}^{\D}\D)=\mathcal{T}^{\D^{*}}\D^{*}.
\end{equation*}

Hence, using the results in Subsection \ref{lasub}, we deduce that

$$(\mathcal{T}^{\D^{*}}\D^{*},\left.\omega_{Q}\right|_{\mathcal{T}^{\D^{*}}\D^{*}\times \mathcal{T}^{\D^{*}}\D^{*}})$$ is a symplectic vector bundle over $\D^{*}$ of rank $2 r$.

As a consequence, there exists a unique section $X_{(g,V,\D)}\in\Gamma(\mathcal{T}^{\D^{*}}\D^{*})$ satisfying
\begin{equation}\label{eqHam}
	\left( i_{X_{(g,V,\D)}} \omega_{Q}|_{\D^{*}} \right)|_{\mathcal{T}^{\D^{*}}\D^{*}}=\left( d H_{(g,V)}|_{\D^{*}} \right)|_{\mathcal{T}^{\D^{*}}\D^{*}} \quad \text{and} \quad X_{(g,V,\D)}\in \mathfrak{X}(\D^{*}).
\end{equation}
Moreover, from \eqref{geometric:nh:symp:eq} and \eqref{eqHam}, we deduce that
\begin{equation}\label{Hamiltonian:correspondence}
	X_{(g,V,\D)}\circ (\flat_{g})|_{\D}=(T\flat_{g})|_{\mathcal{T}^{\D}\D} \circ \Gamma_{(g,V,\D)}.
\end{equation}
So, if $\sigma:I\rightarrow \D^{*}$ is an integral curve of $X_{(g,V,\D)}$ then
\begin{equation*}
	\pi_{Q}\circ \sigma:I\rightarrow Q
\end{equation*}
is a trajectory of the nonholonomic mechanical system $(Q,{(g,V)},\D)$.

\subsection{A contact bundle formulation of the nonholonomic Mau\-per\-tuis-Jacobi principle}

Let $(Q,{(g,V)},\D)$ be a mechanical nonholonomic system and consider the Hamiltonian function $H_{(g,V)}: T^*Q\rightarrow {\mathbb R}$ along with the corresponding Hamiltonian vector field $X_{(g,V,\D)}\in \mathfrak{X}(\D^{*})$.

Suppose that $e\in\R$ is such that $U_{e}=\{ q\in Q \ | \ e>V(q) \}$ is non-empty. Again consider the Jacobi metric $g_{e}$ defined in $U_{e}$ defined on \eqref{jacmet} as well as the distribution $\D_{e}$ and its dual distribution
\begin{equation*}
	\D_{e}^{*}=\bigcup_{q\in U_{e}}\D_{q}^{*}\subseteq T^{*} U_{e}.
\end{equation*}
$\D_{e}^{*}$ is a vector bundle over $U_{e}$ with vector bundle projection $\tau_{e}^{*}:\D_{e}^{*} \rightarrow U_{e}$.

In the Hamiltonian side of the nonholonomic kinetic system $(U_{e},g_{e},\D_{e})$, we will denote by $X_{(g_{e},\D_{e})}\in \mathfrak{X}(\D_{e}^{*})$ the corresponding Hamiltonian vector field.

As we know $$(\mathcal{T}^{\D_{e}^{*}}\D_{e}^{*},\left.\omega_{Q}\right|_{\mathcal{T}^{\D_{e}^{*}}\D_{e}^{*}\times \mathcal{T}^{\D_{e}^{*}}\D_{e}^{*}})$$ is a symplectic vector bundle over $\D_{e}^{*}$ and also
\begin{equation}\label{eqHame}
\left( i_{X_{(g_{e},\D_{e})}} \omega_{Q}|_{\D_{e}^{*}} \right)|_{\mathcal{T}^{\D_{e}^{*}}\D_{e}^{*}}=\left( d H_{g_{e}}|_{\D_{e}^{*}} \right)|_{\mathcal{T}^{\D_{e}^{*}}\D_{e}^{*}},
\end{equation}
where $H_{g_{e}}:T^{*}U_{e}\rightarrow \R$ is the Hamiltonian function in the Hamiltonian side of the kinetic nonholonomic system $(U_{e},g_{e},\D_{e})$. It is important to note that the Hamiltonian function $H_{g_{e}}$ is given by
\begin{equation*}
	H_{g_{e}}(\alpha_{q})=\frac{1}{2}g_{e}^{\sharp}(\alpha_{q},\alpha_{q}),
\end{equation*}
where $g_{e}^{\sharp}$ is the \textit{Jacobi co-metric} which is given by
\begin{equation}\label{Jacobi:cometric}
	g_{e}^{\sharp}=\frac{1}{e-V}g^{\sharp}.
\end{equation}

Let us introduce the subset $S_{e}^{*}$ of $\D_{e}^{*}$ given by
\begin{equation*}
	S_{e}^{*} = \bigcup_{q\in U_{e}} \{ \alpha_{q}\in\D_{q}^{*} \ | \ \|\alpha_{q}\|_{g}^{2}=2(e-V(q)) \}.
\end{equation*}

Then we may prove the following result:
\begin{theorem}[Contact bundle formulation of the nonholonomic Maupertuis-Jacobi principle]\label{maup-noh}
	Using the notation we have introduced before, the following statements hold:
	\begin{enumerate}
		\item The subset $S_{e}^{*}$ satisfies
			\begin{equation*}
				S_{e}^{*} = \left( H_{(g,V)}|_{\D_{e}^{*}} \right)^{-1}(e)= \left( H_{g_{e}}|_{\D_{e}^{*}} \right)^{-1}(1)
			\end{equation*}
			and, in addition, if $\alpha_{q}\in S_{e}^{*}$ then
			\begin{equation*}
				\left( d H_{(g,V)} (\alpha_{q}) \right)|_{\mathcal{T}^{\D_{e}^{*}}\D_{e}^{*}} =\left( d H_{g_{e}}(\alpha_{q}) \right)|_{\mathcal{T}^{\D_{e}^{*}}\D_{e}^{*}}\neq 0,
			\end{equation*}
			so $S_{e}^{*}$ is a submanifold of codimension $1$ in $\D_{e}^{*}$. In fact,
			\begin{equation*}
				\begin{split}
					T_{\alpha_{q}}S_{e}^{*} & = \{ X\in	T_{\alpha_{q}}\D_{e}^{*} \ | \ \langle dH_{(g,V)}(\alpha_{q}),X \rangle = 0 \}  \\
					& = \{ X\in	T_{\alpha_{q}}\D_{e}^{*} \ | \ \langle dH_{g_{e}}(\alpha_{q}),X \rangle = 0 \}
				\end{split}
			\end{equation*}
			and $S_{e}^{*}$ is a bundle over $U_{e}$ with fiber at $q \in U_{e}$ the sphere centred at $0_{q}\in\D_{e}^{*}$ and radius $\sqrt{2(e-V(q))}$, with respect to the Riemannian metric $g$.
			
		\item If $\mathcal{C}_{e}$ is defined by
			\begin{equation*}
				\mathcal{C}_{e} = \bigcup\limits_{\begin{split}
					\alpha_{q} & \in\D_{q}^{*}\\ q & \in U_{e}
					\end{split}} \left( T_{\alpha_{q}}S_{e}^{*}\cap \mathcal{T}_{\alpha_{q}}^{\D_{e}^{*}}\D_{e}^{*} \right)
			\end{equation*}
			then $\mathcal{C}_{e}$ is a vector bundle over $S_{e}^{*}$ which admits a contact bundle structure and the Reeb section $R_{e}$ is just $X_{(g_{e},\D_{e})}|_{\D_{e}^{*}}$.
			\item We have that
				\begin{equation*}
					(e-V)|_{U_{e}}R_{e}=X_{(g,V,\D)}|_{S_{e}^{*}}.
				\end{equation*}
			
			\item If $v_{q}\in\D_{q}$ is a non-zero vector with $q\in U_{e}$ and $c_{{\mathcal P}_{q}(v_{q})}:J\rightarrow U_{e}$, $c_{\mathcal{Q}_{q}(v_{q})}:I\rightarrow U_{e}$ are the nonholonomic trajectories of the systems $(U_{e}(g,V)|_{U_{e}},\D_{e})$, $(U_{e},g_{e},\D_{e})$ with initial velocities ${\mathcal P}_{q}(v_{q})$ and $\mathcal{Q}_{q}(v_{q})$, respectively, then
			\begin{equation*}
				c_{{\mathcal P}_{q}(v_{q})}(s)=c_{\mathcal{Q}_{q}(v_{q})}(h(s)),
			\end{equation*}
			where $h:J\rightarrow I$ is a strictly increasing reparametrization satisfying
			\begin{equation*}
				\frac{d h}{d s}=e-V\circ c_{{\mathcal P}_{q}(v_{q})}, \quad h(0)=0.
			\end{equation*}
	\end{enumerate}
\end{theorem}

\begin{remark}{\rm 
In the above theorem we used some notations introduced in the previous sections, namely the projections $${\mathcal P}_{q}:\D_{q}\setminus\{0_{q}\}\rightarrow \left( E_{{(g,V)}} \right)^{-1}(e)\cap \D_{q}$$ given by
\begin{equation*}
	{\mathcal P}_{q}(v_{q})=\sqrt{2(e-V(q))} \frac{v_{q}}{\|v_{q}\|_{g}}
\end{equation*}
and $\mathcal{Q}_{q}:T_{q}Q\setminus \{0_{q}\}\rightarrow \left( E_{{g_{e}}} \right)^{-1}(1)\cap \D_{q}$ given by
\begin{equation*}
	\mathcal{Q}_{q}(v_{q})=\sqrt{\frac{{2}}{e-V(q)}} \frac{v_{q}}{\|v_{q}\|_{g}}.
\end{equation*}
}
\end{remark}
\begin{proof}
	Let us prove each item in the theorem by order of appearance:
	\begin{enumerate}
		\item We have that
			\begin{equation*}
				\|\alpha_{q}\|_{g}^{2}=2(e-V(q))
			\end{equation*}
			is equivalent to
			\begin{equation*}
					\frac{\|\alpha_{q}\|_{g}^{2}}{2(e-V(q))} = 1
			\end{equation*}
			and so, using the definition of the Jacobi co-metric $g_{e}^{\sharp}$ in \eqref{Jacobi:cometric} we have that
			\begin{equation*}
				\frac{1}{2}\|\alpha_{q}\|_{g_{e}}^{2} = 1,
			\end{equation*}
			which proves that $\alpha_{q} \in \left( H_{(g,V)}|_{\D_{e}^{*}} \right)^{-1}(e)$ if and only if $\alpha_{q} \in \left( H_{g_{e}}|_{\D_{e}^{*}} \right)^{-1}(1)$.
			
			Now, let $\Delta^{*}$ be the Euler vector field of $\D^{*}$ defined by
			\begin{equation*}
				\Delta^{*}(\alpha_{q}) = (\alpha_{q})_{\alpha_{q}}^{V}=\frac{d}{d t}\Big|_{t=0}((1+t)\alpha_{q}) \in \mathcal{T}_{\alpha_{q}}^{\D_{e}^{*}}\D_{e}^{*}.
			\end{equation*}
			Then, if $\alpha_{q} \in S_{e}^{*}$ we have that
			\begin{equation*}
				\left\langle  d H_{(g,V)} (\alpha_{q}) ,\Delta^{*}(\alpha_{q}) \right\rangle = \|\alpha_{q}\|^{2}_{g}=2(e-V(q))>0
			\end{equation*}
			as well as
			\begin{equation*}
				\langle  d H_{g_{e}} (\alpha_{q}), \Delta^{*}(\alpha_{q}) \rangle = \|\alpha_{q}\|^{2}_{g_{e}}=2>0.
			\end{equation*}
			Hence, $S_{e}^{*}$ is a submanifold of $\D_{e}^{*}$ of codimension 1 and
			\begin{equation*}
				\begin{split}
					T_{\alpha_{q}}S_{e}^{*} & =\{ X\in T_{\alpha_{q}}\D_{e}^{*} \ | \ \langle d H_{(g,V)} (\alpha_{q}) ,X \rangle = 0 \} \\
					& = \{ X\in T_{\alpha_{q}}\D_{e}^{*} \ | \ \langle d H_{g_{e}} (\alpha_{q}) ,X \rangle = 0 \}.
				\end{split}
			\end{equation*}
			
			Thus,
			$$T_{\alpha_{q}}\D_{e}^{*} = T_{\alpha_{q}}S_{e}^{*} \oplus \langle \Delta^{*}(\alpha_{q}) \rangle.$$
			Therefore, using that $\Delta^{*}$ is vertical with respect to the projection $\tau_{e}^{*}:\D_{e}^{*}\rightarrow U_{e}$, it follows that the restriction of $\tau_{e}^{*}$ to $S_{e}^{*}$ is also a bundle with projection $\tau_{e}^{*}|_{S_{e}^{*}}:S_{e}^{*}\rightarrow U_{e}$. In addition, it is easy to prove that the fiber of $\tau_{e}^{*}|_{S_{e}^{*}}$ at $q\in U_{e}$ is just the sphere centred at $0_{q}\in\D_{e}^{*}$ and radius $\sqrt{2(e-V(q))}$, with respect to the Riemannian metric $g$.
			
			\item If $\alpha_{q}\in S_{e}^{*}$ then, from the previous item, we deduce that the set
			$$T_{\alpha_{q}}^{\D_{e}^{*}} \D_{e}^{*} \cap T_{\alpha_{q}} S_{e}^{*}$$
			is a vector subspace of codimension $1$ of $\mathcal{T}_{\alpha_{q}}^{\D_{e}^{*}} \D_{e}^{*}$. Therefore, 
			\begin{equation*}
				\mathcal{C}_{e} = \bigcup\limits_{\begin{split}
						\alpha_{q} & \in S_{e}^{*}\\ q & \in U_{e}
				\end{split}} \left( T_{\alpha_{q}}S_{e}^{*}\cap \mathcal{T}_{\alpha_{q}}^{\D_{e}^{*}}\D_{e}^{*} \right)
			\end{equation*}
			 is a vector bundle over $S_{e}^{*}$ with rank $2r -1$ (we recall that $\mathcal{T}_{\alpha_{q}}^{\D_{e}^{*}}\D_{e}^{*}$ is a $2r$-dimensional symplectic vector space).
			
			Now, we consider the sections $(\theta_{Q})_{e}$ and $(\omega_{Q})_{e}$ of the vector bundles $\mathcal{C}_{e}^{*}\rightarrow S_{e}^{*}$ and $\Lambda^{2}\left( \mathcal{C}_{e}^{*} \right)\rightarrow S_{e}^{*}$, respectively, given by
			\begin{equation*}
				(\theta_{Q})_{e}(\alpha_{q})=\frac{1}{2}\theta_{Q}(\alpha_{q})|_{(\mathcal{C}_{e})_{\alpha_{q}}}
			\end{equation*}
			and
			\begin{equation*}
				(\omega_{Q})_{e}(\alpha_{q})=\frac{1}{2}\omega_{Q}(\alpha_{q})|_{(\mathcal{C}_{e})_{\alpha_{q}}\times (\mathcal{C}_{e})_{\alpha_{q}}}
			\end{equation*}
			for $\alpha_{q}\in S_{e}^{*}$.
			
			We will see that $((\theta_{Q})_{e},(\omega_{Q})_{e})$ is a contact bundle structure on the vector bundle $C_{e}\rightarrow S_{e}^{*}$, that is,
			\begin{equation*}
				(\theta_{Q})_{e} \wedge (\omega_{Q})_{e}^{r-1} \in \Gamma ( \Lambda^{2r} (C_{e}^{*}) )
			\end{equation*}
			is non-vanishing at every point of $S_{e}^{*}$. In fact, using that
			\begin{equation*}
				X_{(g_{e},\D_{e})}(\alpha_{q})\left( H_{g_{e}}|_{\D_{e}^{*}} \right)=0
			\end{equation*}
			it follows that $X_{(g_{e},\D_{e})}(\alpha_{q}) \in (\mathcal{C}_{e})_{\alpha_{q}}$. Thus, we deduce that $X_{(g_{e},\D_{e})}|_{S_{e}^{*}} \in \Gamma(\mathcal{C}_{e})$. In addition,
			\begin{equation*}
				\begin{split}
					\langle (\theta_{Q})_{e}(\alpha_{q}),X_{(g_{e},\D_{e})}(\alpha_{q}) \rangle & = \frac{1}{2}\langle \theta_{Q}(\alpha_{q}),X_{(g_{e},\D_{e})}(\alpha_{q}) \rangle \\
					& = \frac{1}{2}\langle \alpha_{q}, T_{\alpha_{q}}\tau_{e}^{*} \left(X_{(g_{e},\D_{e})}(\alpha_{q})\right) \rangle,
				\end{split}
			\end{equation*}
			where we used the definition of the canonical 1-form of the cotangent bundle and $\tau_{e}^{*}:\D_{e}^{*}\rightarrow U_{e}$ is the bundle projection. On the other hand, from \eqref{Hamiltonian:correspondence}, we have that
			\begin{equation*}
				X_{(g_{e},\D_{e})}\circ (\flat_{g_{e}})|_{\D_{e}}=(T\flat_{g_{e}})|_{\mathcal{T}^{\D_{e}}\D_{e}} \circ \Gamma_{(g_{e},\D_{e})}.
			\end{equation*}
			Now, denote by $\sharp_{g_{e}}:\D_{e}^{*}\rightarrow \D_{e}$ the inverse map of the flat isomorphism $\flat_{g_{e}}:\D_{e}\rightarrow \D_{e}^{*}$. Then, using that 
			\begin{equation*}
				T_{\alpha_{q}}\tau_{e}^{*} \circ (T\flat_{g})|_{\mathcal{T}^{\D}\D} = T_{\sharp_{g_{e}}(\alpha_{q})}\tau_{e},
			\end{equation*}
			where $\tau_{e}:\D_{e}\rightarrow U_{e}$ is the canonical bundle projection, we deduce that
			\begin{equation*}
				\langle (\theta_{Q})_{e}(\alpha_{q}),X_{(g_{e},\D_{e})}(\alpha_{q}) \rangle = \frac{1}{2}\langle \alpha_{q}, T_{\sharp_{g_{e}}(\alpha_{q})}\tau_{\D_{e}} \left( \Gamma_{(g_{e},\D_{e})} \circ \sharp_{g_{e}}(\alpha_{q})\right) \rangle.
			\end{equation*}	
			But, since $\Gamma_{(g_{e},\D_{e})}$ is a SODE the previous relation reduces to
			\begin{equation*}
				\begin{split}
					\langle (\theta_{Q})_{e}(\alpha_{q}),X_{(g_{e},\D_{e})}(\alpha_{q}) \rangle & = \frac{1}{2}\langle \alpha_{q}, \sharp_{g_{e}}(\alpha_{q}) \rangle \\
					& = \frac{1}{2} \|\alpha_{q}\|_{g_{e}}^{2}=1.
				\end{split}
			\end{equation*}
			Moreover, we have that
			\begin{equation*}
				\left. \left[ i_{X_{(g_{e},\D_{e})}(\alpha_{q})} (\omega_{Q})_{e}(\alpha_{q}) \right] \right|_{(\mathcal{C}_{e})_{\alpha_{q}}}=d H_{g_{e}}(\alpha_{q})|_{(\mathcal{C}_{e})_{\alpha_{q}}}=0.
			\end{equation*}
			This implies that $((\theta_{Q})_{e},(\omega_{Q})_{e})$ is a contact bundle structure on the vector bundle $\mathcal{C}_{e}$ and that $X_{(g_{e},\D_{e})}|_{S_{e}^{*}}\in \Gamma(C_{e})$ is the Reeb section of this contact structure, that is,
			\begin{equation*}
				i_{X_{(g_{e},\D_{e})}|_{S_{e}^{*}}}(\theta_{Q})_{e}= 1, \quad i_{X_{(g_{e},\D_{e})}|_{S_{e}^{*}}}(\omega_{Q})_{e}= 0.
			\end{equation*}
			\item Using that
			\begin{equation*}
				\left( X_{(g,V,\D)}|_{S_{e}^{*}} \right)\left( H_{(g,V)}|_{\D_{e}^{*}} \right)=0
			\end{equation*}
			it follows that $X_{(g,V,\D)}|_{S_{e}^{*}} \in \Gamma(\mathcal{C}_{e})$. In addition, proceeding as in the previous item, one may prove that if $\alpha_{q}\in S_{e}^{*}$ then
			\begin{equation*}
				\langle (\theta_{Q})_{e}(\alpha_{q}),X_{(g,V,\D)}(\alpha_{q}) \rangle =\frac{1}{2}\|\alpha_{q}\|_{g}^{2}=e-V(q)
			\end{equation*}
			and
			\begin{equation*}
				\left. \left[ i_{X_{(g,V,\D)}}(\alpha_{q}) (\omega_{Q})_{e}(\alpha_{q}) \right] \right|_{(\mathcal{C}_{e})_{\alpha_{q}}}=d H_{(g,V)}(\alpha_{q})|_{(\mathcal{C}_{e})_{\alpha_{q}}}=0.
			\end{equation*}
			Therefore,
			\begin{equation}\label{wer}
				(e-V(q))|_{U_{e}}X_{(g_{e},\D_{e})}|_{S_{e}^{*}}=X_{(g,V,\D)}|_{S_{e}^{*}}.
			\end{equation}
			\item It is easy to prove that the following diagram commutes:
			\begin{figure}[htb!]
				\begin{center}
					\begin{tikzcd}
						& \mathcal{D}_{q}\setminus\{0_{q}\} \arrow[ld, "{\mathcal P}_{q}"'] \arrow[rd, "\mathcal{Q}_{q}"] &                               \\
						E_{{(g,V)}}^{-1}(e)\cap \D_{q} \arrow[rd, "\flat_{g}"'] &                                                                                      & E_{{g_{e}}}^{-1}(1)\cap \D_{q} \arrow[ld, "\flat_{g_{e}}"] \\
						& (S_{e}^{*})_{q}                                                                            &                              
					\end{tikzcd}
				\end{center}
			\caption{Commutative diagram.}\label{diagram:flat:projectors}
			\end{figure}
	
		Thus, if $v_{q}\in \D_{q}\setminus\{0_{q}\}$ then
		\begin{equation}\label{flat:projectors}
			\flat_{g}({\mathcal P}_{q}(v_{q}))=\flat_{g_{e}}(\mathcal{Q}_{q}(v_{q}))=\alpha_{q}\in S_{e}^{*}.
		\end{equation}
		Now, we consider the integral curves $\sigma_{\alpha_{q}}:J\rightarrow S_{e}^{*}$ and $\sigma_{\alpha_{q}}^{e}:I\rightarrow S_{e}^{*}$ (with $0\in I,J$) of the vector fields $X_{(g,V,\D)}|_{S_{e}^{*}}$ and $X_{(g_{e},\D_{e})}|_{S_{e}^{*}}$, respectively, satisfying the initial conditions
		\begin{equation*}
			\sigma_{\alpha_{q}}(0)=\sigma_{\alpha_{q}}^{e}(0)=\alpha_{q}.
		\end{equation*}
		Then, using Equation (\ref{wer}) in the previous item , it follows that there exists a strictly increasing reparametrization $h:J\rightarrow I$ such that
		\begin{equation*}
			\frac{d h}{d s}=e-V\circ \pi_{\D^{*}}\circ \sigma_{\alpha_{q}}, \quad h(0)=0
		\end{equation*}
		and
		\begin{equation*}
			\sigma_{\alpha_{q}}(s)=\sigma_{\alpha_{q}}^{e}(h(s)), \quad \text{for } s\in J,
		\end{equation*}
		with $\pi_{\D^{*}}:\D^{*}\rightarrow Q$ the canonical projection. But, recall that, if $v_{q}=\sharp_{q}(\alpha_{q})$ then using \eqref{Hamiltonian:correspondence}, Figure \ref{diagram:flat:projectors} and \eqref{flat:projectors}, we deduce that
		\begin{equation*}
			\pi_{\D^{*}}\circ \sigma_{\alpha_{q}}=c_{{\mathcal P}_{q}(v_{q})} \quad \text{and} \quad 	\pi_{\D^{*}}\circ \sigma_{\alpha_{q}}^{e}=c_{\mathcal{Q}_{q}(v_{q})},
		\end{equation*}
		which implies the result.
	\end{enumerate}
\end{proof}

%%%%%%%%%%%%%%%%%%%%%%%%%%%%%%%%%%%%%%%%%%%%%%%%%%%%%%%%%%

\begin{remark}{\bf A coordinate derivation of Maupertuis-Jacobi principle (see also \cite{Koiller})}\label{Koiller:remark}
	{\rm 
		Having chosen a system of coordinates $(q^i)$, $1\leq i\leq n=\dim Q$ then we induce  a system of coordinates $(q^i, \dot{q}^i)$ on $TQ$. In these coordinates, the Lagrangian $L_{(g,V)}: TQ\rightarrow {\mathbb R}$ is written as
		\[
		L(q^i, \dot{q}^i)= \frac{1}{2}g_{ij}(q) \dot{q}^i\dot{q}^j-V(q)
		\]
		where $g_{ij}=g(\partial/\partial q^i, \partial/ \partial q^j)$.
		The linear velocity constraints are determined by the distribution $\D$ where $\hbox{rank} \D=m\leq n$ and it is locally determined by its annihilator: 
		\[
		\D^o=\hbox{span}\{\mu^{\alpha}=\mu^{\alpha}_i(q)\,  dq^i; m+1\leq \alpha\leq n\}
		\]
		However in the case of nonholonomic mechanics it can be  better  to adapt the coordinates on the tangent bundle to   the linear velocity constraints and  to the Riemannian metric. To this end, consider a local basis 
		$\{X_a, Y_{\alpha}\}$, $1\leq a\leq m$ and $m+1\leq \alpha\leq n$ of vector fields such that locally 
		\[
		{\mathcal D}_q=\hbox{span}\{X_a(q)\} \quad \hbox{and}\quad {\mathcal D}^{\perp, g}_q=\hbox{span}\{Y_{\alpha}(q)\}\, ,
		\]
		where ${\mathcal D}^{\perp, g}_q$ is the Riemannian-orthogonal to ${\mathcal D}$, i.e.
		\[
		g(X_a, Y_{\alpha})=0\, , \quad 1\leq a\leq m\, \quad \hbox{and} \quad m+1\leq \alpha\leq n\, .
		\]
		Denote by $g_{ab}=g(X_a, X_b)$ and consider the Lie bracket:
		\[
		[X_a, X_b]={\mathcal C}_{ab}^c X_c+{\mathcal C}_{ab}^\alpha Y_{\alpha}
		\]
		Observe that the non-vanishing of some of functions ${\mathcal C}_{ab}^\alpha$ implies the non-integrability of the distribution $\D$. 
		
		Obviously we have that  $T_qQ={\mathcal D}_q\oplus {\mathcal D}^{\perp, g}_q$.
		Therefore,  the adapted  basis $\{X_a, Y_{\alpha}\}$  induces  a new  set of coordinates on the tangent bundle $(q^i, y^a, y^{\alpha})$. 
		Observe that the elements $v_q\in {\mathcal D}_q$ are distinguished by the condition $y^{\alpha}=0$. That is, the nonholonomic constraint are now  $y^{\alpha}=0$  and ${\mathcal D}$ is completely described by coordinates $(q^i, y^a)$.
		
		Denote by $\{X^a, Y^{\alpha}\}$ the dual basis corresponding to $\{X_a, Y_{\alpha}\}$ inducing coordinates $(q^i, p_a, p_{\alpha})$ on $T^*Q$ and $(q^i, p_a)$ on $D^*$. 
		The Hamiltonian is now 
		\begin{equation*}
			H_{(g, V)}|_{\D^*}(q^i, p_a)=\frac{1}{2}g^{ab}(q) p_ap_b+V(q) \, . \label{hamilt}
		\end{equation*}
		The equations of motion of a nonholonomic system are written in the system of adapted coordinates $(q^i, p_a)$ as follows (see, for instance, \cite{CodeMaMa,CeFaHoMa}): 
		\begin{subequations}
			%\label{H}
			\begin{align}
				\dot{q}^i&=X^i_b\frac{\partial H_{(g, V)}|_{\D^*}}{\partial p_b}=X^i_b g^{ab}p_a\; , \label{Hm1} \\
				\dot{p}_a&=-{\mathcal C}_{ab}^c p_c\frac{\partial H_{(g, V)}|_{\D^*}}{\partial p_b}-X^i_a\frac{\partial H_{(g, V)}|_{\D^*}}{\partial q^i}\; \\
				&=-{\mathcal C}_{ab}^c g^{bd}p_cp_d-X^i_a\left( \frac{1}{2}\frac{\partial g^{cb}}{\partial q^i}p_cp_b+\frac{\partial V}{\partial q^i}\right),\  \label{Hm2}
			\end{align}
		\end{subequations}
		where $X_{a}=X^i_a \frac{\partial}{\partial q^{i}}$. The dynamics is precisely the given by the vector field $X_{(g,V, \D)}$ intrinsically defined  in Equation (\ref{eqHam}). 
		
		From the other hand,  if we consider the Hamiltonian $H_{g_e}|_{\D_e^*}:\D_e^*\rightarrow {\mathbb R} $: 
		\begin{equation*}
			H_{g_e}|_{\D^*_e}(q^i, p_a)=\frac{1}{2(e-V(q))}g^{ab}p_ap_b \, . \label{hamilt-1}
		\end{equation*}
		Then the corresponding nonholonomic equations are: 
		\begin{subequations}
			\label{H}
			\begin{align}
				\dot{q}^i=&\frac{1}{e-V(q)} X^i_b g^{ab}p_a\; , \label{H1} \\
				\dot{p}_a
				=&-\frac{1}{e-V(q)}{\mathcal C}_{ab}^c g^{bd}p_cp_d\\
				&-X^i_a(q)\left( \frac{1}{2(e-V(q))}\frac{\partial g^{cb}}{\partial q^i}p_cp_b+\frac{1}{2(e-V(q))^2}\frac{\partial V}{\partial q^i}g^{cb}p_cp_b\right)\  \label{H2}
			\end{align}
		\end{subequations}
		These equations are precisely the ones defined by the integral curves of the vector field $X_{(g_e,  \D_e)}$ given in Equation (\ref{eqHame}). 
		
		Therefore
		\[
		X_{(g_e,  \D_e)}-\frac{1}{e-V(q)} X_{(g,V, \D)}|_{D_e^*}=X^i_a(q)\left(\frac{1}{2(e-V(q))}\frac{\partial V}{\partial q^i}g^{cb}p_cp_b-\frac{\partial V}{\partial q^i}\right)\frac{\partial}{\partial p_a}
		\]
		
		Along the set $S_{e}^{*} = \left( H_{(g,V)}|_{\D_{e}^{*}} \right)^{-1}(e)$  we have that
		$\frac{1}{2}g^{cb}p_cp_b=e-V(q)$ and in consequence, 
		\[
		{\mathcal R}_e=X_{(g_e, \D_e)}|_{S_{e}^{*} }=\frac{1}{e-V(q)} X_{(g, V, \D)}|_{S_{e}^{*} }
		\]
		as appears in Theorem \ref{maup-noh}. 
	}
\end{remark}

\section{Proof of the main Theorem \ref{mechanical:nonholonomic:geodesic:theorem}}\label{section5}

Now we have all the ingredients to   prove of Theorem \ref{mechanical:nonholonomic:geodesic:theorem}
since it is a direct consequence combining first  the  nonholonomic Maupertuis-Jacobi principle stated in Theorem \ref{maup-noh} and then Theorem \ref{kin-theorem}. We just add a few reasons why we take the open ball $B_{g}\left(0_{q};\sqrt{\frac{{2 \varepsilon}}{e-V(q)}}\right)$, with $\varepsilon$ a sufficiently small positive number, as the domain of the map $\text{exp}_{q}^{nh,e}$:
\begin{itemize}
	\item It is clear that $B_{g}\left(0_{q};\sqrt{\frac{{2 \varepsilon}}{e-V(q)}}\right)$ is a star-shaped open subset of $\D_{q}$ about $0_{q} \in \D_{q}$;
	\item If $v_{q} \in \D_{q} \setminus \{0_{q}\}$, then $\mathcal{Q}_{q}(v_{q})\in B_{g}\left(0_{q};\sqrt{\frac{{2}}{e-V(q)}}\right)$. So, if we fix $\varepsilon>0$ small enough, it is possible to choose a sufficiently small positive number $\lambda$ such that
	\begin{equation*}
		h(s)\mathcal{Q}_{q}(v_{q}) \in B_{g}\left(0_{q};\sqrt{\frac{{2 \varepsilon}}{e-V(q)}}\right), \ \forall s \in [0,\lambda]
	\end{equation*}
	(note that $h(0)=0$);
	\item Using the previous facts, we can directly apply Theorem \ref{kin-theorem} to the map $\text{exp}_{q}^{nh,e}:B_{g}\left(0_{q};\sqrt{\frac{{2 \varepsilon}}{e-V(q)}}\right) \subseteq \D_{q} \rightarrow Q$.
\end{itemize}

\section{Example}

\begin{example}
	Let us first consider a mechanical nonholonomic system describing a particle with unitary mass in euclidean three dimensional space $Q=\R^{3}$ equipped with the euclidean metric $g$, subjected to a  potential force $V:Q\rightarrow \R$ given by
	$$V(x,y,z)= z,$$
	and to the nonholonomic constraint determined by $$\D = \{ (q,\dot{q})\in TQ \ | \ \dot{z}=y\dot{x} \}.$$
	
	Let $e\in R$ be a fixed energy value and consider the set
	$$U_{e}=\{ (x,y,z)\in Q \ | \ z<e \}$$
	where the Jacobi metric $$g_{e}=(e-z)g$$
	is defined.
	The kinetic nonholonomic system $(g_{e},\D_{e})$ associated to the mechanical nonholonomic system $(g,V,\D)$ is associated to the kinetic Lagrangian $L_{g_{e}}:TU_{e}\rightarrow \R$ given by
	\begin{equation*}
	L_{g_{e}}(q,\dot{q})=\frac{e-z}{2}\left( \dot{x}^{2} + \dot{y}^{2} + \dot{z}^{2} \right).
	\end{equation*}
	To observe explicitly the results of Theorem \ref{maup-noh}, it is easier to work on the Hamiltonian side and using a basis adapted to $\D$, as in Remark \ref{Koiller:remark}. In that sense, we will use the basis
	given by
	$$X_{1}=\frac{\partial}{\partial x} + y \frac{\partial}{\partial z}, \quad X_{2} = \frac{\partial}{\partial y}$$
	spanning $\D$ and the vector
	$$Y_{1}=-y\frac{\partial}{\partial x} + \frac{\partial}{\partial z}$$
	spanning the orthogonal complement $\D^{\bot}$. Hence, we obtain the following non-vanishing components of the Riemannian metric
	$$g_{11}=1+y^2, \quad g_{22}=1.$$
	Finally, the non-vanishing structure functions $(C_{a b}^{c})$ relative to this basis are
	$$C_{1 2}^{1} = -\frac{y}{y^2+1}=-C_{2 3}^{3}, \quad C_{2 3}^{1}=-\frac{1}{y^2+1}=C_{1 2}^{3}.$$
	
	The Hamiltonian function is written with respect to this basis as
	$$H_{(g,V)}|_{\D^{*}}(q^{i},p_{a})= \frac{1}{2} \left( \frac{p_{1}^{2}}{y^{2}+1} + p_{2}^{2} \right) +z$$
	and the corresponding Hamiltonian equations in this adapted coordinates are
	\begin{equation*}
		\begin{cases}
			\dot{x}= \frac{p_{1}}{y^{2}+1}\\
			\dot{y}= p_{2}\\
			\dot{z}= \frac{y p_{1}}{y^{2}+1}\\
		\end{cases}
		\quad
		\begin{cases}
			\dot{p}_{1} = \frac{y p_{1} p_{2}}{y^{2}+1} - y\\
			\dot{p}_{2} = 0 \\
		\end{cases}
	\end{equation*}
	
	On the other hand, the kinetic Hamiltonian function $H_{g_{e}}:T^{*}U_{e}\rightarrow \R$ is given on these coordinates by
	\begin{equation*}
	H_{g_{e}}(q^{i},p_{a})=\frac{1}{2(e-z)}\left( \frac{p_{1}^{2}}{y^{2}+1} + p_{2}^{2} \right)
	\end{equation*}
	implying the following Hamiltonian equations
	\begin{equation*}
		\begin{cases}
			\dot{x}= \frac{1}{e-z} \frac{p_{1}}{y^{2}+1}\\
			\dot{y}= \frac{1}{e-z} p_{2}\\
			\dot{z}= \frac{1}{e-z} \frac{y p_{1}}{y^{2}+1}\\
		\end{cases}
		\quad
		\begin{cases}
			\dot{p}_{1} = \frac{1}{e-z} \frac{y p_{1} p_{2}}{y^{2}+1} - \frac{y}{2(e-z)^2}\left( \frac{p_{1}^{2}}{y^{2}+1} + p_{2}^{2} \right)\\
			\dot{p}_{2} = 0. \\
		\end{cases}
	\end{equation*}
	Then it is clear that if we restrict to the set $S^{*}_{e}=\left( H_{(g,V)}|_{\D^{*}_{e}} \right)^{-1}(e)$, we have that
	$$\frac{1}{2(e-z)}\left( \frac{p_{1}^{2}}{y^{2}+1} + p_{2}^{2} \right) = e -z$$
	on $S^{*}_{e}$, showing that
	\begin{equation*}
		X_{(g_{e},\D_{e})}|S_e^*=\frac{1}{e-z}X_{(g, V,\D)}|S_e^*.
	\end{equation*}
	It is now clear that the integral curves of $X_{(g, V,\D)}$ must be a reparametrization of the integral curves of the Hamiltonian vector field $X_{(g_{e},\D_{e})}$ 	on $S^{*}_{e}$.
\end{example}

\begin{example}
	The vertical rolling disk with harmonic potential in the steering angle. Consider the mechanical Lagrangian function $L_{(g,V)}:TQ\rightarrow \R$ in the manifold $Q=\R^{2}\times \Es^{1}\times \Es^{1}$ given by
	\begin{equation*}
	L_{(g,V)}(q,\dot{q})=\frac{1}{2}(\dot{x}^{2}+\dot{y}^{2}+\dot{\theta}^{2}+\dot{\varphi}^{2})-\frac{\varphi^{2}}{2},
	\end{equation*}
	subject to the constraint
	$$\D = \{ (q,\dot{q})\in TQ \ | \ \dot{x}=\dot{\theta} \cos \varphi, \ \dot{y}= \dot{\theta} \sin \varphi \}.$$
	It is not difficult to show that the general solution is
	\begin{equation*}
	\begin{cases}
	x(t) & = \int_{0}^{t} \cos(\varphi(s)) \ ds + x_{0} \\
	y(t) & = \int_{0}^{t} \sin(\varphi(s)) \ ds + y_{0} \\
	\theta (t) & = \Omega t + \theta_{0} \\
	\varphi (t) & = \varphi_{0} \cos(t)+\omega \sin(t),
	\end{cases}
	\end{equation*}
	with $q_{0}=(x_{0},y_{0},\theta_{0},\varphi_{0})\in Q$, $(\Omega,\omega)\in \R^{2}$ a coordinate chart on $\D_{q_{0}}$, representing the initial angular velocities. Then the nonholonomic exponential map, which is the map $\text{exp}_{q_{0}}^{nh}:\D_{q_{0}}\rightarrow Q$ given by
	$$\text{exp}_{q_{0}}^{nh}(v_{q_{0}})=(\tau_{Q}\circ \phi_{1}^{(g,V,\D)})(v_{q}),$$
	where $\phi_{t}^{(g,V)}$ is the flow of the nonholonomic mechanical system $(L_{(g,V)},\D)$, is a local diffeomorphism onto its image and so its inverse map is $R_{q_{0}}^{nh}:\M_{q_{0}}^{nh}\rightarrow \D_{q_{0}}$ given by
	\begin{equation*}
	R_{q_{0}}^{nh}(\theta,\varphi)=\left( \theta-\theta_{0},\frac{\varphi-\varphi_{0} \cos(1)}{\sin(1)} \right).
	\end{equation*}
	
	The corresponding kinetic nonholonomic system is determined by the Lagrangian function $L:{g_{e}}:TU_{e}\rightarrow \R$ given by
	\begin{equation*}
	L_{g_{e}}(q,\dot{q})=\frac{e-\frac{\varphi^{2}}{2}}{2}(\dot{x}^{2}+\dot{y}^{2}+\dot{\theta}^{2}+\dot{\varphi}^{2}).
	\end{equation*}
	After some computations, we may eliminate the Lagrange multipliers appearing in Lagrange-d'Alembert equations and find that the trajectories of the nonholonomic system $(L_{g_{e}},\D_{e})$ must satisfy
	\begin{equation*}
	\begin{cases}
	\dot{x} & = \dot{\theta} \cos \varphi \\
	\dot{y} & = \dot{\theta} \sin \varphi \\
	\ddot{\theta} & = \frac{2 \varphi \dot{\varphi}\dot{\theta}}{e-\frac{\varphi^{2}}{2}} \\
	\ddot{\varphi} & = \frac{\varphi \dot{\varphi}^2-\varphi \dot{\theta}^2 }{e-\frac{\varphi^{2}}{2}}.
	\end{cases}
	\end{equation*}
	Then the trajectories of this system form the exponential map $\text{exp}_{q_{0}}^{(g_{e},\D_{e})}:\D_{q_{0}}\rightarrow Q$.
	
	Moreover, using Theorem \ref{nhm}, we know there is a strictly increasing function  $h:J\rightarrow I$ satisfying
	\begin{equation*}
	\frac{d h}{d s}=e-V\circ c_{P_{q}(v_{q})}, \quad h(0)=0.
	\end{equation*}
	Solving the differential equations, we obtain that
	\begin{equation*}
		h(s)=es-\frac{1}{2}\left( \frac{(\varphi_{0}^{2}-\omega^{2})\cos s \sin s}{2}+\frac{(\varphi_{0}^{2}+\omega^{2})s}{2}+\varphi_{0}\omega\sin^{2}s \right).
	\end{equation*}
\end{example}

\begin{example}
	The vertical rolling disk with linear potential in the steering angle. Consider the mechanical Lagrangian function $L_{(g,V)}:TQ\rightarrow \R$ in the manifold $Q=\R^{2}\times \Es^{1}\times \Es^{1}$ given by
	\begin{equation*}
	L_{(g,V)}(q,\dot{q})=\frac{1}{2}(\dot{x}^{2}+\dot{y}^{2}+\dot{\theta}^{2}+\dot{\varphi}^{2})-\varphi,
	\end{equation*}
	subject to the constraint
	$$\D = \{ (q,\dot{q})\in TQ \ | \ \dot{x}=\dot{\theta} \cos \varphi, \ \dot{y}= \dot{\theta} \sin \varphi \}.$$
	It is not difficult to show that the general solution is
	\begin{equation*}
	\begin{cases}
	x(t) & = \int_{0}^{t} \cos(\varphi(s)) \ ds + x_{0} \\
	y(t) & = \int_{0}^{t} \sin(\varphi(s)) \ ds + y_{0} \\
	\theta (t) & = \Omega t + \theta_{0} \\
	\varphi (t) & = \omega t + \varphi_{0}-\frac{t^2}{2},
	\end{cases}
	\end{equation*}
	with $q_{0}=(x_{0},y_{0},\theta_{0},\varphi_{0})\in Q$, $(\Omega,\omega)\in \R^{2}$ a coordinate chart on $\D_{q_{0}}$, representing the initial angular velocities. Then the nonholonomic exponential map $\text{exp}_{q_{0}}^{nh}:\D_{q_{0}}\rightarrow Q$ given by
	$$\text{exp}_{q_{0}}^{nh}(v_{q_{0}})=(\tau_{Q}\circ \phi_{1}^{(g,V,\D)})(v_{q}),$$
	where $\phi_{t}^{(g,V, \D)}$ is the flow of the nonholonomic mechanical system $(L_{(g,V)},\D)$, is a local diffeomorphism onto its image and so its inverse map is $R_{q_{0}}^{nh}:\M_{q_{0}}^{nh}\rightarrow \D_{q_{0}}$ given by
	\begin{equation*}
	R_{q_{0}}^{nh}(\theta,\varphi)=\left( \theta-\theta_{0},\varphi-\varphi_{0}+\frac{1}{2} \right).
	\end{equation*}
	
	The corresponding kinetic nonholonomic system is determined by the Lagrangian function $L_{g_{e}}:TU_{e}\rightarrow \R$ given by
	\begin{equation*}
	L_{g_{e}}(q,\dot{q})=\frac{e-\varphi}{2}(\dot{x}^{2}+\dot{y}^{2}+\dot{\theta}^{2}+\dot{\varphi}^{2}).
	\end{equation*}
	After some computations, we may eliminate the Lagrange multipliers appearing in the corresponding Lagrange-d'Alembert equations and find that the trajectories of the nonholonomic system $(L_{g_{e}},\D_{e})$ must satisfy
	\begin{equation*}
	\begin{cases}
	\dot{x} & = \dot{\theta} \cos \varphi \\
	\dot{y} & = \dot{\theta} \sin \varphi \\
	\ddot{\theta} & = \frac{ \dot{\varphi}(\dot{\theta}+\sin(\varphi)\dot{y}+\cos(\varphi)\dot{x})}{2e-2\varphi} \\
	\ddot{\varphi} & = \frac{\dot{\varphi}^2-2 \dot{\theta}^2 }{2e-2\varphi}.
	\end{cases}
	\end{equation*}
	Then the trajectories of this system form the exponential map $\text{exp}_{q_{0}}^{(g_{e},\D_{e})}:\D_{q_{0}}\rightarrow Q$.
	
	Moreover, using Theorem \ref{nhm}, we know there is a strictly increasing function  $h:J\rightarrow I$ satisfying
	\begin{equation*}
	\frac{d h}{d s}=e-V\circ c_{P_{q}(v_{q})}, \quad h(0)=0.
	\end{equation*}
	Solving the differential equations, we obtain that
	\begin{equation*}
	h(s)=es+\frac{s^{3}}{6}-\frac{\omega s^{2}}{2}-\varphi_{0}s.
	\end{equation*}
%The inverse function of $h$ is the function
%\begin{equation*}
%\begin{split}
%h^{-1}(t)= & \left(\omega^3+3\omega\varphi_{0}-3\omega e+3t + \right.\\
% &\left. +\sqrt{3 e^2 \omega^2+6 e \omega^2 \varphi_{0}+6 \omega^3 t-3 \omega^2 \varphi_{0}^2+8 e^3-24 e^2 \varphi_{0}-18 e \omega t+24 e \varphi_{0}^2+18 \omega t \varphi_{0}-8 \varphi_{0}^3+9 t^2}\right)^{1/3}\\
% & -(-\omega^2-2 \varphi_{0}+2 e)/ \\
% & \left(\omega^3+3 \omega \varphi_{0}-3 \omega e+3 t + \right. \\
% &\left. + \sqrt{3 e^2 \omega^2+6 e \omega^2 \varphi_{0}+6 \omega^3 t-3 \omega^2 \varphi_{0}^2+8 e^3-24 e^2 \varphi_{0}-18 e \omega t+24 e \varphi_{0}^2+18 \omega t \varphi_{0}-8 \varphi_{0}^3+9 t^2}\right)^{1/3}\\
% &+\omega
%\end{split}
%\end{equation*}
	Moreover, by the definition of nonholonomic exponential map we have that
	$$\text{exp}_{q}^{nh,e}(v_{q})=c^{e}_{v_{q}}(1),$$
	where $c^{e}_{v_{q}}$ is the trajectory of the kinetic nonholonomic system $(L_{g_{e}},\D_{e})$. In addition, note that every non-zero vector in $\D$ might be uniquely written in the form
	$$v_{q}=\lambda(v_{q})\mathcal{Q}_q(v_{q}), \quad \lambda(v_{q})=\sqrt{\frac{e-V(q)}{2}}\|v_{q}\|_{g}.$$
	Hence, by the homothetic property of kinetic nonholonomic trajectories we deduce
	$$\text{exp}_{q}^{nh,e}(v_{q})=c^{e}_{\mathcal{Q}_q(v_{q})}(\lambda(v_{q})).$$
	Alternatively, using again Theorem \ref{nhm} we may also write
	$$\text{exp}_{q}^{nh,e}(v_{q})=c_{\mathcal{P}_q(v_{q})}(h^{-1}(\lambda(v_{q}))).$$
	
	Let $(\Omega,\omega)$ be coordinates on $\D$ associated to the basis
	$$\{\cos \varphi \frac{\partial}{\partial x}+\sin \varphi \frac{\partial}{\partial y}+\frac{\partial}{\partial \theta},\frac{\partial}{\partial \varphi} \}$$
	and $(\theta,\varphi)$ coordinates on $\M^{nh,e}_{q_{0}}=\M^{nh}_{q_{0}}$ under which 
	$$c_{v_{q}}(t)=\left( \Omega t + \theta_{0},\omega t + \varphi_{0}-\frac{t^{2}}{2} \right).$$
	Now, $E_{(g,V)}(\Omega,\omega)=e$ if and only if the initial velocity $\Omega$ is equal to
	$$\Omega^{\pm}:=\pm \sqrt{(e-\varphi_{0})-\frac{\omega}{2}},$$
	so that
	$$c_{P(v_{q})}(t)=\left( \Omega^{\pm} t + \theta_{0},\omega t + \varphi_{0}-\frac{t^{2}}{2} \right).$$
	If $k=h^{-1}\circ \lambda$ then we have that
	$$\text{exp}_{q}^{nh,e}(\Omega^{\pm},\omega)=\left( \Omega^{\pm} k(\Omega^{\pm},\omega) + \theta_{0},\omega k(\Omega^{\pm},\omega) + \varphi_{0}-\frac{k^{2}(\Omega^{\pm},\omega)}{2} \right).$$
	Considering the flat metric in $\D$ as a Gauss metric, i.e., the metric
	$$\mathcal{G}_{0}=d\Omega\otimes d\Omega+d\omega \otimes d\omega$$
	then the reparametrization by the function $h$ of the unit energy geodesics with respect to the metric
	$$g_{q}^{nh,e}=\left((\text{exp}_{q}^{nh,e})^{-1}\right)^{*}\mathcal{G}_{0}$$
	are just the mechanical nonholonomic trajectories with energy $e$ with initial point $q$. Therefore, the nonholonomic trajectories are reparametrizations of minimizing geodesics for the Riemannian metric $g_{q}^{nh,e}$. In particular, they minimize the Riemannian length associated with this metric.
\end{example}

%%%%%%%%%%%%%%%%%%%%%%%%%%%%%%%%%%%%%%%%%%%%%%%%%%%%%%%%%%%%%%%%%%%%%%%%%%%%%%%%%%%%%%%%%%%%%%%%%%%%%%%%%%%%%%%%%%%%%%%%%%%%%%%%%%%%%%%%%%%%%%%%%%%%%%%%%%%%%%%%%%%%%%%%%%%%%%%%%%%%%%%%%%%%%%%%%%%%%%%%%%%%%%%%%%%%%%%%%%%%%%%%%%%%%%%%%%%%%%%%%%%%%%%%%%%%%%%%%%%%%%%%%%%%%%%%%%%%%%%%%%%%%%%%%%%%%%%%%%%%%%%%%%%%%%%%%%%%%%%%%%%%%%%%%%%%%%%%%%%%%%%%%%%%%%%%%%%%%%%%%%%%%%%%%%%%%%%%%%%%%%%%%%%%%%

%For acknowledgements section, please don't number the section, please begin it with \section*{Acknowledgements}
\section*{Acknowledgments} 
D. Mart{\'\i}n de Diego and A. Simoes  acknowledge financial support from the Spanish Ministry of Science and Innovation, under grants PID2019-106715GB-C21, MTM2016-76702-P,  the “Severo Ochoa Programme for Centres of Excellence" in R\&D  (CEX2019-000904-S) and from the Spanish National Research Council, through the ``Ayuda extraordinaria a Centros de Excelencia Severo Ochoa'' (20205\-CEX001). A. Simoes is supported by the FCT (Portugal) research fellowship SFRH/BD/129882/2017, partially funded by the European Union (ESF). J.C. Marrero  acknowledges the partial support by European Union (Feder) grant PGC2018-098265-B-C32.

%%%%%%%%%%%%%%%%%%%%%%%%%%%%%%%%%%%%%%%%%%%%%%%%%%%%%%%%%%%%%%%%%%%%%%%%%%%%%%%%%%%%%%%%%%%%%%%%%%%%%%%%%%%%%%%%%%%%%%%%%%%%%%%%%%%%%%%%%%%%%%%%%%%%%%%%%%%%%%%%%%%%%%%%%%%%%%%%%%%%%%%%%%%%%%%%%%%%%%%%%%
%\bibliographystyle{apacite}
\bibliography{thesisreferences}{}

\end{document}